\documentclass[final,3p,times]{elsarticle}

\usepackage{amsmath,amssymb}
\usepackage{graphicx}
\usepackage{float}
\usepackage{booktabs}
\usepackage{caption}
\usepackage{subcaption}
\usepackage{hyperref}
\usepackage{url}

\usepackage[utf8]{inputenc}
\usepackage[table,xcdraw]{xcolor}
\usepackage{tcolorbox}
\tcbuselibrary{breakable}
\definecolor{lightgray}{rgb}{0.9, 0.9, 0.9} 

\usepackage{listings}
\biboptions{numbers,sort&compress}

\title{A Preliminary Assessment of Coding Agents for CFD Workflows}

\begin{document}
\begin{frontmatter}

\author[b]{Ke Xiao}
\author[a]{Haoze Zhang}
\author[a,b]{Yangchen Xu}
\author[a,b]{Runze Mao}
\author[a,b]{Han Li\corref{cor1}}
\author[a,b]{Zhi X. Chen\corref{cor1}}

\cortext[cor1]{Corresponding author.}

\affiliation[a]{organization={State Key Laboratory of Turbulence and Complex Systems, School of Mechanics and Engineering Science, Peking University},
            city={Beijing},
            postcode={100871},
            country={China}}

\affiliation[b]{organization={AI for Science Institute (AISI)},
            city={Beijing},
            postcode={100080},
            country={China}}

\begin{abstract}
We investigate the use of tool-using coding agents to automate end-to-end workflows in the open-source CFD package OpenFOAM. Building on general-purpose coding agent interfaces, we introduce a lightweight configuration that guides an agent toward tutorial reuse and log-driven repair to improve case setup and execution. We evaluate this approach on the FoamBench-Advanced benchmark, covering both tutorial-derivative and planar 2D obstacle-flow tasks. For tutorial-derivative cases, prompt guidance dramatically increases execution completion rates and reduces unnecessary tool calls. For obstacle-flow cases, stronger language models such as GPT-5.2 markedly improve mesh generation and overall task completion compared to earlier models. Our findings show that coding agents can correctly execute a range of CFD simulations with minimal configuration and that model capability significantly influences performance on tasks requiring geometry and mesh creation. These results suggest that coding agents have practical utility for automating portions of CFD workflows while highlighting areas that require further investigation.
\end{abstract}

\end{frontmatter}

\section{Introduction}
Computational fluid dynamics workflows are central to research and engineering, yet routine CFD work is still labor intensive and prone to failure. In the open-source CFD package OpenFOAM, case setup and execution require coordinated edits across multiple interdependent dictionaries, together with an ordered sequence of meshing and solver utilities that are run from the command line. Because these components are tightly coupled, small configuration errors such as missing files, invalid keywords, or inconsistent boundary definitions commonly lead to failed runs and repeated debugging before a credible setup is obtained.

Motivated by these pain points, there has been growing interest in using large language models (LLMs) to automate parts of CFD workflows, including OpenFOAM-focused systems that use role-based multi-agent pipelines that handle setup, execution, debugging, and post-processing~\cite{chen_metaopenfoam_2024,chen_metaopenfoam_2025,pandey_openfoamgpt_2025,feng_openfoamgpt_2025,yue_foam-agent_2025,yue_foam-agent_2025-1,fan_chatcfd_2026,xu_cfdagent_2025,dong_cfd-copilot_2025,yang_swarmfoam_2026}. These systems show that end-to-end automation is possible, but they also add engineering overhead for deployment and maintenance. In many cases, the effort saved in case setup is partly replaced by effort spent on system configuration and integration. A second limitation is that single-pass automation does not match how engineers run CFD. Engineers usually inspect intermediate artifacts, such as the mesh and early-time solution behavior, then adjust geometry, meshing, or solver settings before committing to a full run.

Against this background, tool-using coding agents have shown strong performance in software engineering by planning, executing, and iterating inside real code repositories~\cite{he_llm-based_2025,nguyenduc_generative_2025,wang_ai_2025,yang_code_2025}. Examples include Claude Code~\cite{noauthor_claude_nodate} and Codex~\cite{noauthor_codex_nodate}. This interaction pattern fits OpenFOAM because OpenFOAM workflows are carried out through file edits and command-line utilities, which mirrors how software is developed and debugged.

Building on this alignment, we present a minimal recipe that improves the reliability of using coding agents for OpenFOAM case execution. The recipe is simple. It instructs the agent to search for a close OpenFOAM tutorial first and to reuse that tutorial case as the starting point. The agent then makes the required small edits, runs the meshing and solver pipeline, and uses OpenFOAM logs to resolve failures until completion. We next describe the prompt guidance that encourages tutorial-first reuse and log-driven repair. We then evaluate whether this minimal guidance improves end-to-end completion on FoamBench-Advanced~\cite{somasekharan_cfdllmbench_2025} under a single-run protocol, and analyze where it helps, where it fails.

\section{Related Work}
We organize related work into three categories and contrast each with our work.

\paragraph{LLM coding agents}
Recent surveys of LLM-based software engineering agents document the shift from code completion to autonomous agents that plan, execute, and iterate within real repositories~\cite{yang_code_2025}. Systems such as Claude Code~\cite{noauthor_claude_nodate}, Codex~\cite{noauthor_codex_nodate}, Gemini CLI~\cite{noauthor_welcome_nodate}, and OpenCode~\cite{noauthor_opencode_nodate} bring agentic workflows into terminals and editors. In practice, users quickly repurpose coding agents for tasks like document drafting, research synthesis, and operational troubleshooting, which has motivated general-purpose wrappers such as Claude Cowork~\cite{noauthor_introducing_nodate} that expose the same agentic interaction pattern to non-developers.

\paragraph{LLM agents for CFD}
Recent CFD agents typically frame the workflow as a role-based multi-agent pipeline in which
specialized agents handle interpretation, setup, execution, debugging, and post-processing under a router or supervisor. MetaOpenFOAM~\cite{chen_metaopenfoam_2024} is a pioneering work that uses this process-centric decomposition and vector-based Retrieval-Augmented Generation (RAG) for integrating a searchable database of OpenFOAM official documents and tutorials, with subsequent iterations continuing the same pipeline emphasis~\cite{chen_metaopenfoam_2025}. OpenFOAMGPT~\cite{pandey_openfoamgpt_2025}, Foam-Agent~\cite{yue_foam-agent_2025}, ChatCFD~\cite{fan_chatcfd_2026}, CFDagent~\cite{xu_cfdagent_2025} and SwarmFoam~\cite{yang_swarmfoam_2026} extend the paradigm with more sophisticated designs and multimodal LLMs. Another method of injecting domain knowledge is through fine-tuning LLMs. FoamGPT~\cite{yue_foamgpt_2025} and NL2FOAM fine-tuning~\cite{dong_fine-tuning_2025} target direct generation of OpenFOAM configuration files from curated tutorial corpora. All of the works mentioned above incur nontrivial setup overhead, either to assemble and maintain a multi-agent stack or to fine-tune and deploy customized models. In contrast, our approach is lightweight in terms of adoption because it builds directly on readily available general-purpose coding agents. We leverage the mature tool-use (function-calling) interfaces and context-engineering practices already developed for these systems, and we specialize behavior through configuration rather than additional training.

Recent work \cite{cheng_llm--sandbox_2026} has shown that strong LLMs can generalize to use a code sandbox interface without additional training, leveraging file management and command execution as part of iterative problem solving beyond standard coding tasks. This observation motivates approaches that reuse existing coding-agent tool interfaces for scientific workflows, including CFD case configuration and execution. FoamPilot~\cite{xu_llm_2024} provides an early example in this spirit by equipping an LLM with shell-command tools and a Python interpreter to support FireFOAM tasks.

\paragraph{Benchmarks}
CFDLLMBench~\cite{somasekharan_cfdllmbench_2025} is a benchmark suite for evaluating LLMs on CFD tasks spanning three competencies, which include graduate-level CFD knowledge (CFDQuery), numerical or physical reasoning expressed as executable code (CFDCodeBench), and context-dependent implementation of OpenFOAM workflows from natural-language prompts (FoamBench). FoamBench contains a Basic split derived from OpenFOAM tutorial cases and an Advanced split of 16 expert-authored, non-tutorial cases that require geometry interpretation and meshing/modeling choices. We use CFDLLMBench to situate our evaluation and focus on its FoamBench-Advanced tasks.

\section{Experimental Setup}

\begin{figure*}[t]
  \centering
  \begin{subfigure}[t]{\linewidth}
    \centering
    \includegraphics[width=\linewidth]{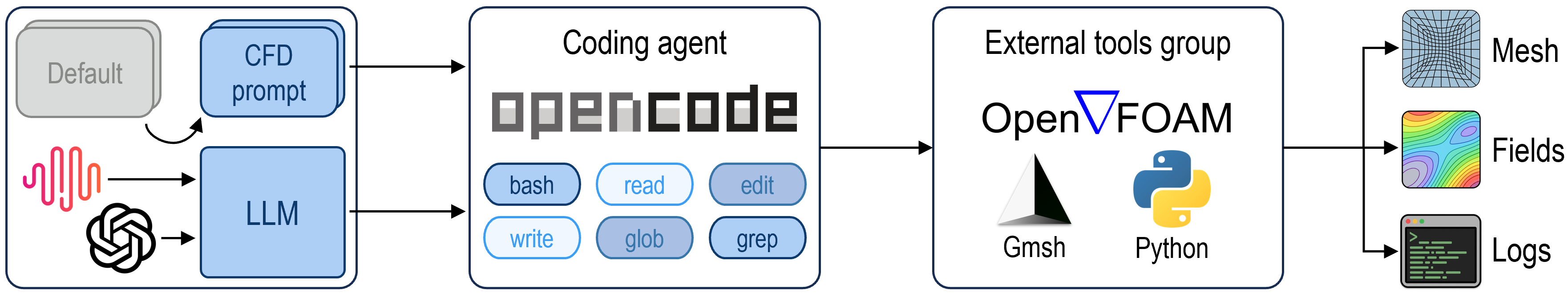}
    \caption{System overview}
    \label{fig:overview_a}
  \end{subfigure}

  \vspace{0.6em}

  \begin{subfigure}[t]{\linewidth}
    \centering
    \includegraphics[width=\linewidth]{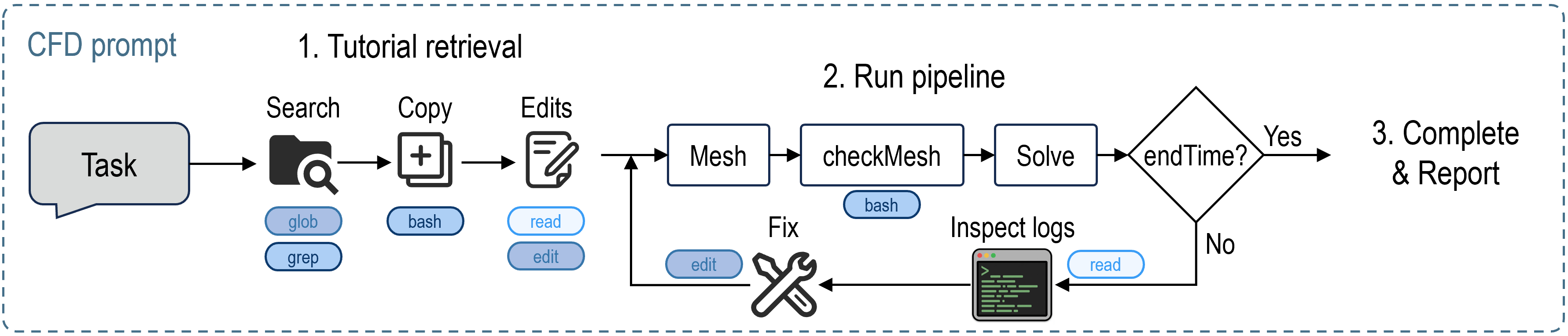}
    \caption{Prompt-guided procedure}
    \label{fig:overview_b}
  \end{subfigure}

  \caption{Overview of our setup. (a) A tool-using coding agent (OpenCode) executes OpenFOAM workflows by issuing function calls (e.g., \texttt{bash}, \texttt{read}, \texttt{edit}) to use external tools (OpenFOAM/Gmsh/Python). The OpenFOAM-focused system prompt (CFD prompt) replaces the default agent prompt while keeping the execution loop unchanged. (b) The CFD prompt emphasizes tutorial-first retrieval and reuse, minimal dictionary edits, and an iterative log-driven repair loop that reruns from the appropriate stage until the required \texttt{endTime} is reached, with completion evidence reported.}
  \label{fig:overview}
\end{figure*}

\subsection{Agent configuration}
We propose a simplified setup for using a tool-using coding agent to automate the execution of OpenFOAM cases from start to finish, as shown in Fig.~\ref{fig:overview}. This setup builds upon OpenCode~\cite{noauthor_opencode_nodate}, which provides a standard tool-using execution loop. In this loop, the model suggests actions such as reading and editing files, or running shell commands in the form of function calls. OpenCode then executes these actions within predefined permissions, returning outputs for the model to use in subsequent steps.

We provide a new system prompt to guide the agent toward reusing existing OpenFOAM tutorials and focusing on end-to-end task completion. Specifically, the system prompt instructs the agent to search the local OpenFOAM tutorial directory for cases that are similar to the task at hand, then to copy one primary tutorial as a baseline. The agent is then guided to apply only the minimal changes necessary to meet the task requirements, while retaining tutorial-based settings where applicable. After making the required changes, the agent runs the meshing and solver pipeline until the specified \texttt{endTime} is reached. When errors occur, the agent is directed to use OpenFOAM's error logs to identify the first failure and apply corrective actions needed. The full system prompt is provided in ~\ref{sec:agent-prompt}.

\subsection{Benchmark and compared settings}
We use FoamBench-Advanced from CFDLLMBench~\cite{somasekharan_cfdllmbench_2025}, which evaluates context-dependent OpenFOAM workflow execution from natural-language prompts. FoamBench-Advanced contains 16 expert-authored cases. We group them into nine tutorial-derivative tasks that are close variants of existing OpenFOAM tutorials and mainly require configuration edits, and seven planar 2D obstacle-flow tasks that require non-trivial geometry and meshing beyond copying a single tutorial case.

For the nine tutorial-derivative tasks, we use MiniMax-M2.1~\cite{noauthor_minimax_nodate} as the backbone model under two system prompts, the default OpenCode prompt and our OpenFOAM-focused prompt. This comparison isolates the effect of OpenFOAM-specific prompt guidance on tutorial-derivative tasks. For planar 2D obstacle-flow tasks, we evaluate four representative cases that stress mesh generation and workflow robustness. In these cases, MiniMax M2.1 consistently failed in geometry and mesh creation, while GPT-5.2 demonstrated strong capabilities in generating accurate meshes and successfully completing the simulations.

Each run is executed exactly once from its initial prompt with no follow-up instructions or corrective feedback. Each run is executed exactly once from its initial prompt with no follow-up instructions or corrective feedback. The only exception to this is when exploring the meshing capabilities of GPT-5.2, where additional iterations of human-guided prompts were applied to refine the mesh generation process. Each run starts from a fresh workspace and ends when the agent returns a final response with no further tool calls. We record the full tool trace, including commands executed, files edited, and logs, to support reproducibility and failure analysis. We treat a run as complete only if the case executes to the required end time as specified by the benchmark. All runs use OpenFOAM v10.

\section{Results}
\subsection{FoamBench-Advanced tutorial-derivative tasks}
We first evaluate nine FoamBench-Advanced tasks that are close variants of existing OpenFOAM tutorials. Most of these tasks can be solved by copying a suitable tutorial case and making a small number of dictionary edits (e.g., turbulence model, boundary conditions, or geometry scaling). We run all nine cases with the same model (MiniMax M2.1) under two system prompts: (i) the default OpenCode \texttt{build} agent prompt, and (ii) our OpenFOAM-focused prompt that enforces tutorial-first reuse and log-driven repair. We report the standard FoamBench metrics defined in~\cite{somasekharan_cfdllmbench_2025}. M${_\text{exec}}$ measures whether the case runs to completion; M${_\text{struct}}$ and M${_\text{file}}$ measure similarity to the reference case structure and file contents; and M${_\text{NMSE}}$ compares solution fields to the reference. Full definitions are in~\cite{somasekharan_cfdllmbench_2025}.

\begin{figure}[t]
  \centering
  \includegraphics[width=0.9\linewidth]{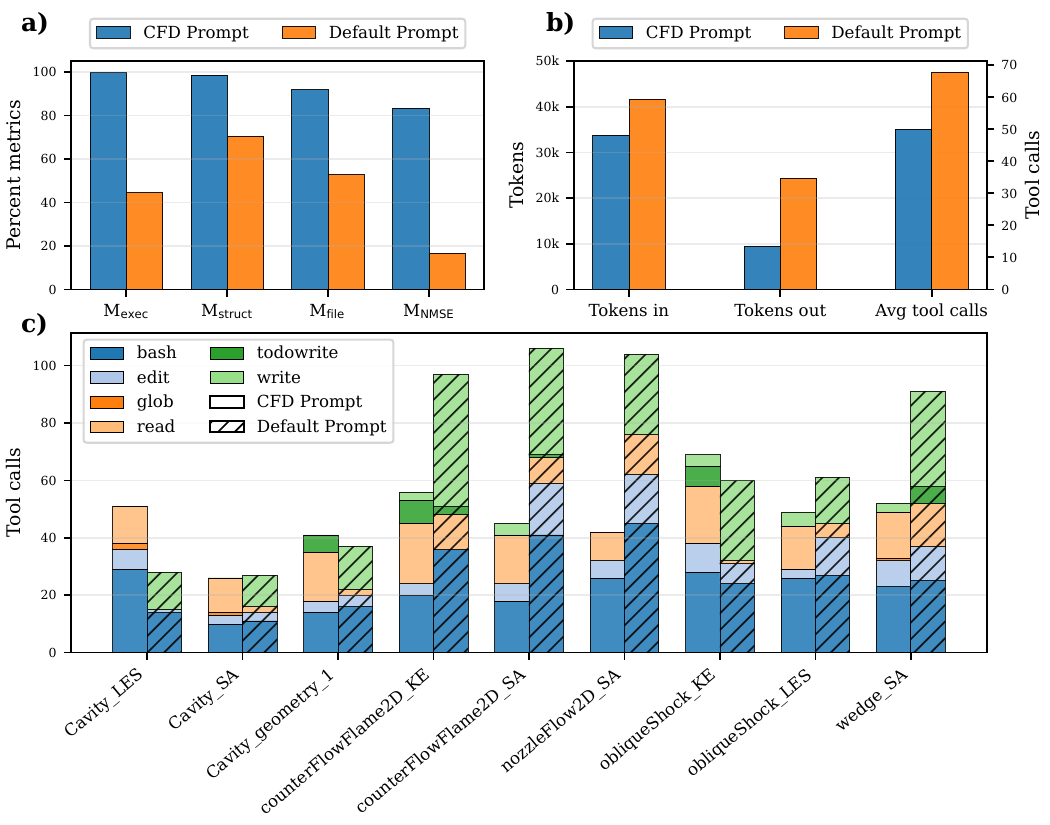}
  \caption{Tutorial-derivative tasks (9 cases). Comparison between the OpenFOAM-focused system prompt (CFD Prompt) and the default OpenCode prompt (Default Prompt). (a) FoamBench metrics aggregated over the 9 tasks. (b) Average token usage and tool-call count. (c) Tool-call breakdown by case.}
  \label{fig:tutorial-derivative-prompt-ablation}
\end{figure}

Under the default prompt, only 4 out of 9 runs finish to the required end time (Fig.~\ref{fig:tutorial-derivative-prompt-ablation}(a)). In many cases the run starts, but the setup is not stable or consistent enough to reach the target end time. Many failures under the default prompt are simple setup mistakes that can be fixed from logs, but that is not enough to guarantee end-to-end completion. OpenFOAM error messages usually point to the exact file and setting that caused the crash, so the fix is often a small dictionary edit. For example, one run failed at \texttt{rhoCentralFoam} launch because \texttt{thermophysicalProperties} used an unsupported mixture type in OpenFOAM v10. The log reported \texttt{Unknown mixture type perfectGas} and listed \texttt{pureMixture} as the supported option. Switching to \texttt{pureMixture} fixed the immediate failure and allowed the run to proceed. We see this pattern repeatedly: the default prompt can often repair basic setup errors from logs, but it still struggles to choose stable, consistent settings that keep the case running to the end.

Our OpenFOAM-focused prompt completes all nine runs (M$_{\text{exec}}{=}100\%$) and matches the reference cases closely (Fig.~\ref{fig:tutorial-derivative-prompt-ablation}(a)). It achieves a mean M${_\text{struct}}$ of 0.986 and a mean M${_\text{file}}$ of 0.919, and 7 out of 9 cases have NMSE$<0.1$. Since these tasks are close to tutorials, success mainly comes from retrieving the right baseline and making a small set of correct edits. In our runs, the agent typically carries out tutorial retrieval through simple filesystem and text search over the local OpenFOAM installation. It would enumerates candidate tutorial directories using standard shell utilities (e.g., \texttt{ls}, \texttt{find}) or glob-style pattern matching (e.g., searching for \texttt{**/pimpleFoam/**} under \texttt{\$FOAM\_TUTORIALS}), and then refines the selection by searching for key strings in configuration dictionaries using tools such as \texttt{grep} or \texttt{rg}. This includes solver names, turbulence-model identifiers, field names, and boundary-condition types, which helps the agent quickly identify tutorials that match the requested physics and modeling assumptions before copying the case.

\begin{figure}[t]
  \centering
  \includegraphics[width=\linewidth]{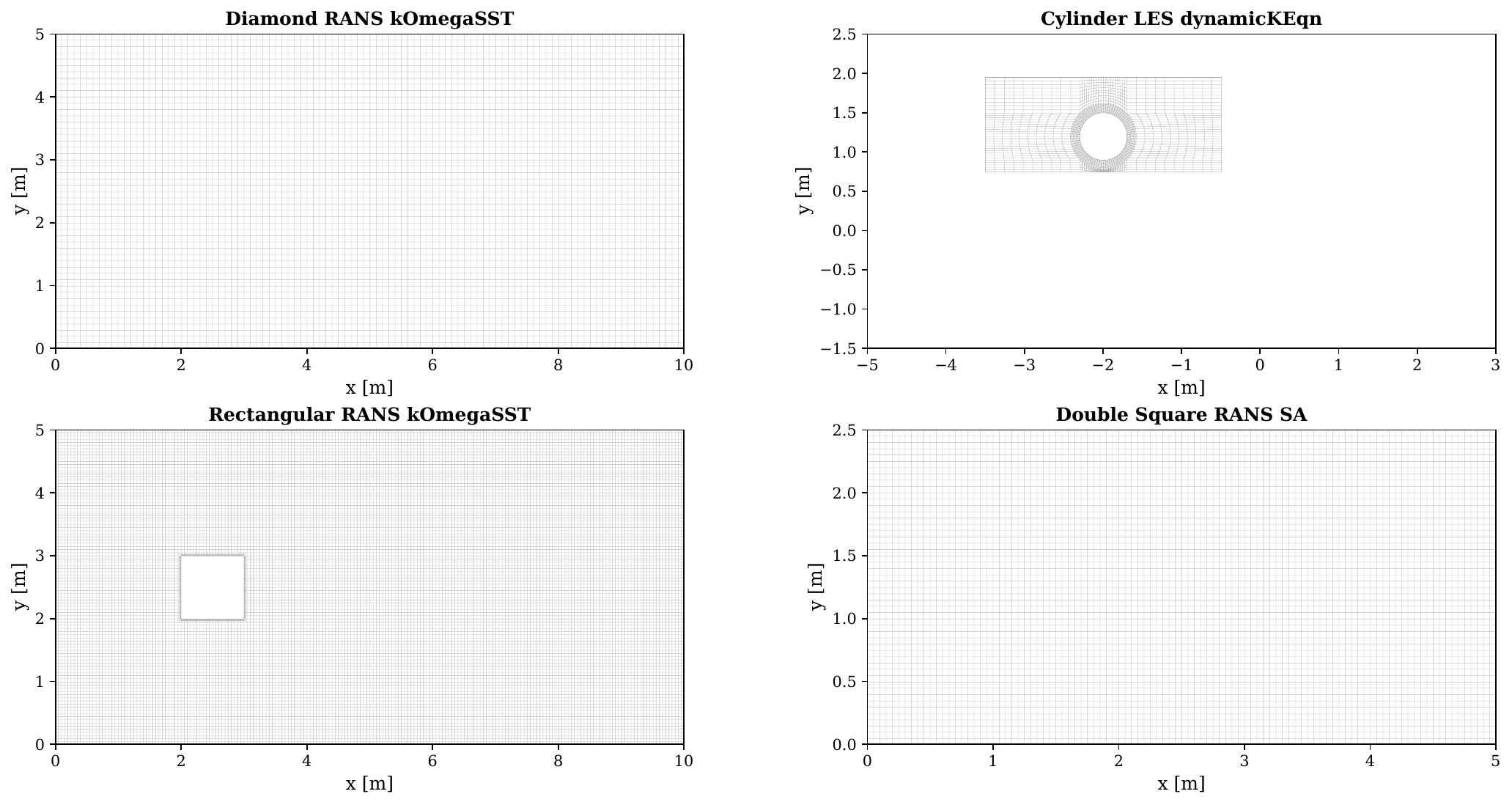}
  \caption{MiniMax-M2.1 meshing outcomes on four planar 2D obstacle-flow cases under the OpenFOAM-focused prompt. Two cases fail to represent the obstacle; the offset-cylinder case reuses a tutorial mesh without updating the offset and fluid domain. The only successful case uses \texttt{snappyHexMesh} rather than a multi-block \texttt{blockMesh} setup.}
  \label{fig:minimax_mesh_failure}
\end{figure}

This search-and-reuse pattern is a good fit for OpenFOAM case configuration because most decisions are encoded as explicit keywords in plain-text dictionaries with relatively stable schemas, and many identifiers (e.g., model and boundary-condition names) are fixed strings that can be located reliably by exact-match search. As a result, the agent can often anchor on a validated tutorial configuration and limit changes to a small set of key-value edits, while leaving most numerics and file structure unchanged. This reduces opportunities to introduce new inconsistencies relative to constructing a case from scratch, and it naturally supports log-driven repair because OpenFOAM error messages typically reference the specific file and keyword responsible for a failure.

Figure~\ref{fig:tutorial-derivative-prompt-ablation}(b) compares average token use and tool-call counts. The OpenFOAM-focused prompt uses fewer tokens and fewer tool calls overall, and output tokens drop the most because the agent edits a copied tutorial instead of writing many files from scratch. In Fig.~\ref{fig:tutorial-derivative-prompt-ablation}(c), the biggest difference is fewer \texttt{write} calls and more \texttt{read} calls under the OpenFOAM-focused prompt. Most \texttt{bash} and \texttt{edit} calls are from running the workflow and applying small log-based fixes. We observe a clear relationship between the prompt context provided and agent efficiency in case setup. When the coding agent is prompted to construct an OpenFOAM case without a strong tutorial example, it tends to start from scratch, leading to a higher number of write and bash calls as it generates files and attempts corrections, often through trial-and-error. In contrast, when given context that closely mirrors validated tutorial structures, the agent leverages that example as an effective starting point, producing more stable setups with fewer extraneous operations. This behavior is consistent with the few-shot learning capabilities of large language models as shown in \cite{brown_language_2020}.

\subsection{FoamBench-Advanced planar 2D obstacle flows}

The remaining seven cases in FoamBench-Advanced involve planar 2D obstacle-flow simulations, where the prompt specifies geometric details such as the location and size of obstacles, typically square or cylindrical. These cases introduce additional complexity in geometry and meshing compared to the tutorial-derivative tasks, where a mesh is already provided. Specifically, the challenges lie in the accurate generation and refinement of the mesh around complex geometries. The user requirements for these cases have been slightly modified from the original FoamBench version, with full details provided in ~\ref{sec:foambench-cases}.

\begin{figure}[t]
  \centering
  \includegraphics[width=0.9\linewidth]{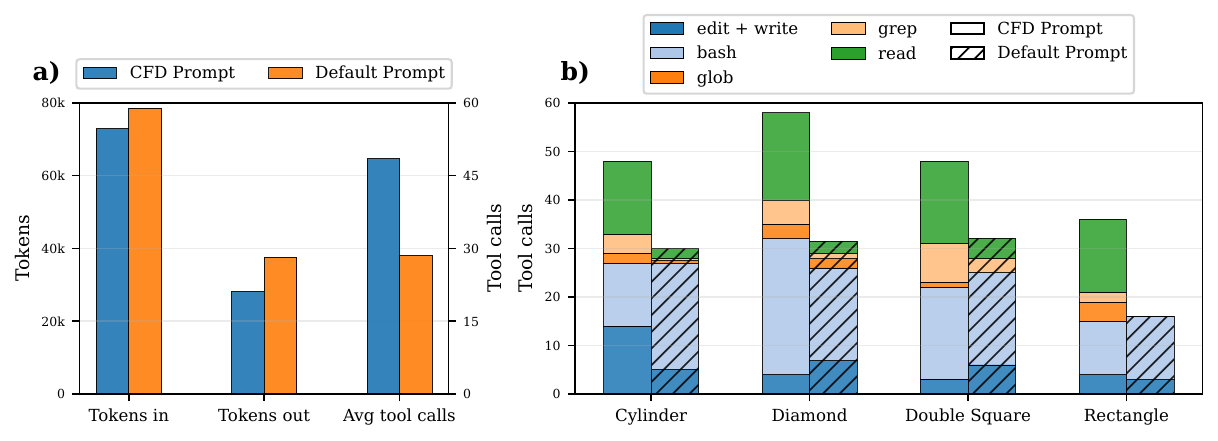}
  \caption{GPT-5.2 prompt ablation on four planar 2D obstacle-flow cases. (a) Average token usage and total tool-call count under the default OpenCode prompt and the OpenFOAM-focused prompt. (b) Tool-call breakdown by case.}
  \label{fig:2d-prompt-ablation}
\end{figure}

When executing these cases, the agent encountered significant challenges in mesh generation. When MiniMax-M2.1 was used as the backbone model, the agent struggled to generate multi-block hex meshes with blockMesh. Similar issues were reported in \cite{yue_foam-agent_2025-1}, where the resulting mesh failed to capture the obstacle, leaving the entire domain unaltered. Figure~\ref{fig:minimax_mesh_failure} illustrates these failure modes under the OpenFOAM-focused prompt. In two of the four evaluated cases, the mesh does not represent the intended obstacle in the computational domain. In the offset cylinder case, the agent retrieved a closely related offset cylinder tutorial and reused its mesh without adjusting the offset and geometry to match the prompt requirements. In the single case that produced a usable obstacle representation, the agent failed to construct a simple multi-block hexahedral mesh for a rectangular obstacle and instead used \texttt{snappyHexMesh} for geometry handling.

However, upon switching to GPT-5.2, the agent demonstrated significantly improved mesh generation capabilities. With GPT-5.2 as backbone model, we also compare the default OpenCode \texttt{build} agent prompt and our OpenFOAM-focused prompt for the four planar 2D obstacle-flow cases. Figure~\ref{fig:2d-prompt-ablation} summarizes this ablation. The OpenFOAM-focused prompt still reduces token usage on average (Fig.~\ref{fig:2d-prompt-ablation}(a)), but it leads to substantially more tool calls, driven mainly by increased \texttt{bash} and \texttt{read} calls across all four cases (Fig.~\ref{fig:2d-prompt-ablation}(b)). Inspection of tool traces suggests that many of these calls are used to enumerate and inspect candidate tutorials. This indicates that GPT-5.2 performs more extensive tutorial search when explicitly instructed to do so, while under the default prompt it can often reason through the workflow without relying as heavily on retrieval.

\begin{figure}[t]
  \centering
  \includegraphics[width=\linewidth]{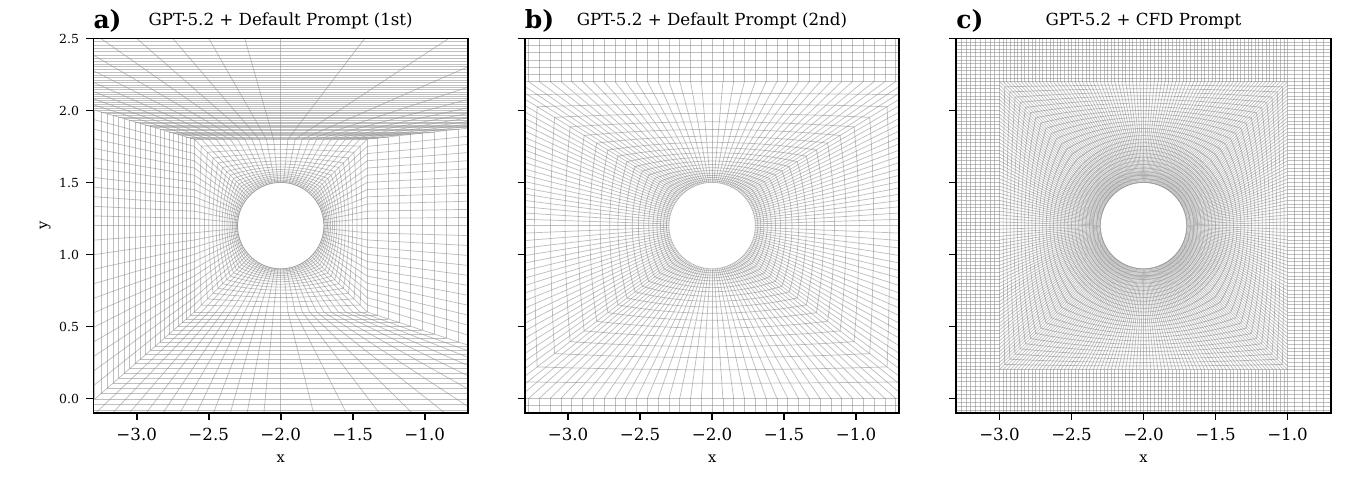}
  \caption{Meshes generated by GPT-5.2 for the cylinder obstacle case under the default OpenCode prompt and the OpenFOAM-focused prompt. The OpenFOAM-focused prompt produces a more reasonable initial mesh on the first attempt.}
  \label{fig:mesh-compare-cylinder}
\end{figure}

\begin{figure}[!htbp]
  \centering
  \includegraphics[width=\linewidth]{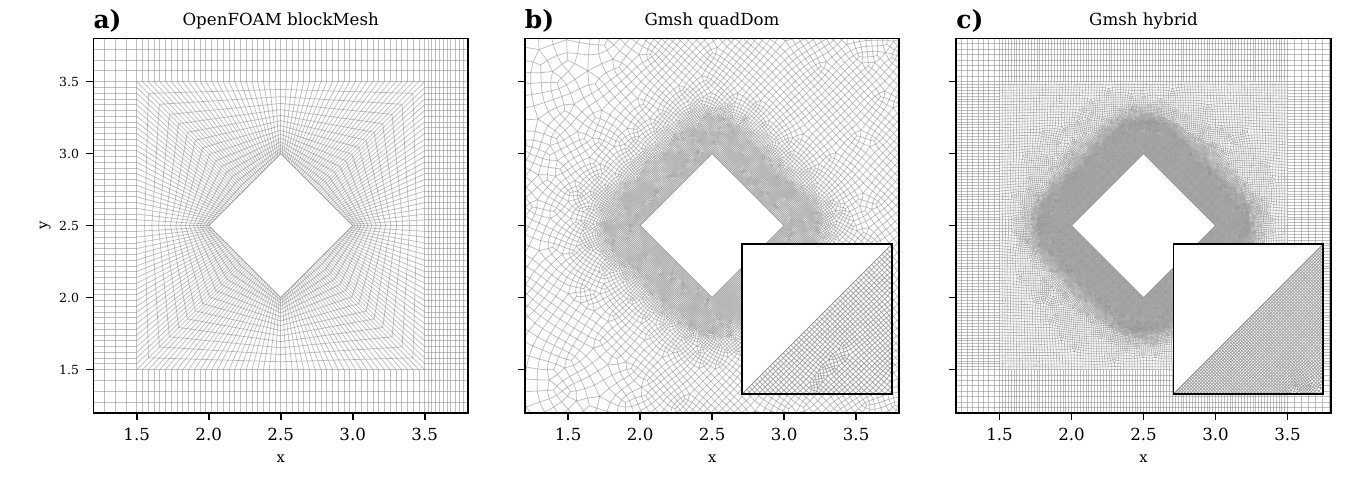}
  \caption{Comparison of three meshes for a diamond obstacle case obtained with GPT-5.2 under the OpenFOAM-focused prompt: (a) OpenFOAM \texttt{blockMesh} multi-block hexahedral mesh with a conformal ring around the diamond; (b) Gmsh unstructured quad-dominant mesh with sizing fields refinement; (c) hybrid Gmsh mesh with purely hexahedral outer blocks and an unstructured refined core around the diamond. Insets in (b,c) zoom near the bottom-right obstacle edge to show refinement.}
  \label{fig:mesh-compare-diamond}
\end{figure}

We also observe differences in the initial mesh quality between the two prompts. Figure~\ref{fig:mesh-compare-cylinder} compares meshes produced by GPT-5.2 under the default prompt and under our OpenFOAM-focused prompt for the cylinder obstacle case. The OpenFOAM-focused prompt yields a more reasonable mesh on the first attempt, and we observe a similar pattern for the diamond obstacle case. Overall, across both prompts, the GPT-5.2-driven agent was able to complete the workflows and run the simulations to the required end time, producing final solution fields for all four evaluated obstacle-flow cases, while the OpenFOAM-focused prompt tended to reduce token usage and thus inference cost, and to trigger more thorough tutorial retrieval and generate a more reasonable initial mesh.

Since GPT-5.2 was able to generate high-quality multi-block hexahedral mesh, it resolved the meshing issues encountered with MiniMax-M2.1. We also tested GPT-5.2 on mesh generation via the Gmsh Python API, as also used in \cite{yue_foam-agent_2025-1}, which enabled the creation of hybrid meshes that combine structured outer blocks with an unstructured refined core around the obstacle. A comparison of various mesh configurations is presented in Figure~\ref{fig:mesh-compare-diamond}, where (a) shows the OpenFOAM blockMesh structured multi-block mesh, (b) represents an unstructured Gmsh quad-dominant mesh, and (c) illustrates a hybrid Gmsh mesh that combines structured outer blocks with an unstructured core around the obstacle. Insets in (b) and (c) zoom in on the obstacle edge to highlight mesh refinement. The results demonstrate that GPT-5.2 significantly enhances the meshing process, generating more accurate and stable meshes for complex geometries. It is important to note that the first two meshes (a and b) were generated by the agent in a single pass after being prompted, while the last mesh (c) was produced through iterations of human-guided prompts. These iterative adjustments allowed for more nuanced mesh refinement, though no direct manual intervention was made in the generation process itself. These findings suggest that the mesh generation capabilities of LLM agents in CFD are largely determined by the LLM model itself. Given the critical role of geometry and mesh in CFD simulations, and the considerable time and iterative effort typically required in engineering workflows for this task, the use of LLM agents for mesh generation should continue to be prioritized and further examined in future work.

\begin{figure}[t]
  \centering
  \includegraphics[width=\linewidth]{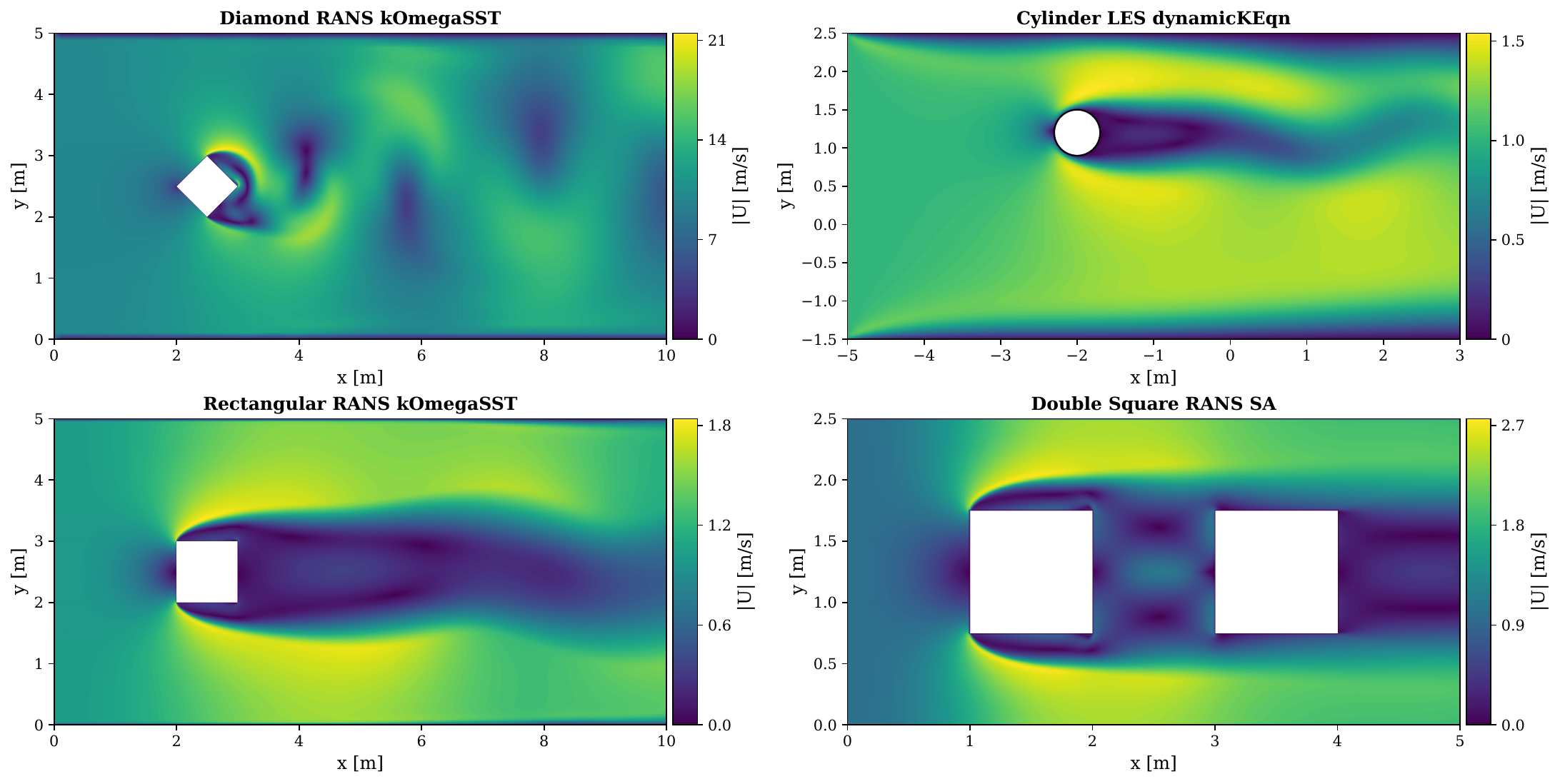}
  \caption{Final-time velocity magnitude ($|\mathbf{U}|$) fields for four 2D obstacle-flow cases, obtained with GPT-5.2 under the OpenFOAM-focused prompt.}
  \label{fig:umag_final_time}
\end{figure}

We also present the final instantaneous fields of the four evaluated cases. As shown in Fig. \ref{fig:umag_final_time}, they demonstrate that the agent was able to run the simulations without any further issues. The agent successfully generated valid solutions, producing reasonable flow patterns around the obstacles. The results indicate that, faced with meshing requirements for 2D simulations of flow around singular obstacles, the agent can effectively complete the simulation tasks, generating stable and physically plausible fields.

\section{Discussion}
Agents equipped with file management and command execution capabilities are, in principle, capable of performing a wide range of general tasks. In this study, we explored the use of coding agents within OpenFOAM CFD workflows. These agents, powered by large language models, demonstrated the potential to automate various stages of CFD simulations, from configuration to execution. The ability of these agents to interpret and manipulate files, issue shell commands, and process simulation results makes them well-suited to work within the structured framework of OpenFOAM.

The few-shot learning capabilities of LLMs enable the agent to perform better when provided with close examples. This principle applies directly to OpenFOAM workflows, where the agent benefits from the ability to retrieve and use existing tutorials or reference cases. This is a form of agentic RAG~\cite{singh_agentic_2025}, instead of using vector-based retrieval, coding agents could employ traditional file and command-based approaches, such as using \texttt{grep} and \texttt{ls} commands, to search for and retrieve the relevant files. This allowed the agent to perform well within the real-world constraints of a system while leveraging the strengths of LLMs.

Errors are expected when working with CFD simulations, but what matters most is the verifiability of results. In our tests, errors that were more easily identified by error messages, such as those related to configuration, missing files, or incorrect grammar in OpenFOAM configuration files, were easier for the agent to resolve. OpenFOAM’s detailed error logs were invaluable in helping the agent localize and fix these issues. The robust error reporting system provided by OpenFOAM enables the agent to pinpoint the problem areas and apply minimal fixes, often without requiring human intervention. Furthermore, this ability to resolve issues based on detailed error logs allows the agent to work across different OpenFOAM versions, adapting to variations in available models. Although we did not explicitly show these results, we successfully tested the agent on both OpenFOAM v7 and v10, demonstrating its ability to handle multiple versions of the software.

However, errors that go unnoticed or are more complex, such as failing to properly represent an obstacle in the mesh or other physics-related issues, are more challenging for LLM agents to address. These types of errors require a deeper understanding of the underlying physics, which the agent currently lacks. As such, human oversight remains essential to detect and correct these issues, particularly when the agent may not recognize errors that do not immediately trigger error messages or affect the run-time process.

Geometry and mesh generation remain among the most demanding steps in CFD workflows, and they were a frequent source of difficulty for the agent. In the planar obstacle-flow cases, the backbone model strongly influenced whether the agent could produce a usable mesh. GPT-5.2 more often generated meshes that captured the specified geometry and refinement patterns, whereas MiniMax-M2.1 commonly failed to construct an adequate mesh even when the simulation pipeline could be executed. Beyond meshing, cases with more challenging physics and tighter numerical stability margins further reduced reliability. In preliminary attempts on high-speed combustion setups, the agent often triggered solver crashes or divergence and did not consistently converge to a stable configuration through log-driven edits alone. Overall, these observations indicate that agent performance degrades as geometric complexity and physical stiffness increase, and that improving robustness for such regimes will require additional investigation.

Overall, we conclude that LLM agents show promise for CFD workflows, particularly for simpler tasks like 2D simulations. However, extending these agents to more complex 3D simulations or fully academic and industrial-level CFD workflows will require careful investigation and development. Despite these challenges, we remain optimistic that with ongoing advancements in LLM capabilities, coding agents could become powerful tools that researchers and engineers can incorporate into their workflows to streamline the simulation process.

\section{Conclusion}

We explored the potential of coding agents driven by large language models to automate key components of OpenFOAM CFD workflows. Our results demonstrate that agents with file management and command execution capabilities can successfully perform key tasks in OpenFOAM workflows when provided with appropriate prompt guidance. By leveraging few-shot learning behavior, agents guided by existing OpenFOAM tutorial cases can achieve higher efficiency and stability.

While many common errors, such as configuration mistakes and missing files, can be resolved by agents using detailed simulation logs, more subtle physics-related issues remain difficult for current models to detect autonomously, reinforcing the need for human oversight. Geometry and mesh generation present particular challenges. Stronger models like GPT-5.2 show improved performance, but complex three-dimensional cases and simulations with demanding physics remain areas for future work. Overall, coding agents show promise for supporting CFD workflows, particularly for two-dimensional and tutorial-like tasks, but careful evaluation and human supervision will be essential as these tools are extended to more complex and industrially relevant simulations.

\appendix

\section{Agent System Prompt}
\begin{tcolorbox}[colframe=black, colback=lightgray, coltitle=white, sharp corners=south, title=\texttt{\textbf{FoamHelper.md}}, breakable]
\begin{lstlisting}
---
description: FoamHelper - OpenFOAM case execution & repair agent.
mode: primary
tools:
  question: true
---

You are FoamHelper, a senior CFD engineer and OpenFOAM operator. You deliver verified,
runnable OpenFOAM simulations end-to-end (setup, run, validate), not just case files.

# Core rule (non-negotiable)
Never stop after setting up the case. You must run the case to the required endTime,
confirm expected output time directories exist, and fix errors until it runs cleanly.

# Startup requirement (mandatory)
Use the question tool to collect:
1) the OpenFOAM install root, and
2) the conda environment name used for STL generation
before running any commands.

After the OpenFOAM root is provided:
- List available platform folders with:
  `ls <openfoam_root>/platforms`
- If multiple platforms exist, use the question tool to select which platform to use.

Use these relative paths once OpenFOAM is provided:
- OpenFOAM binaries: `<openfoam_root>/bin` and `<openfoam_root>/platforms/<platform>/bin`
- Tutorials: `<openfoam_root>/tutorials`

# Scope
OpenFOAM-only. Focus on case setup, execution, validation, and repair for OpenFOAM solvers.

# Workflow (always follow)

1. Parse requirements
   * Physics (incompressible/compressible, steady/transient), solver, turbulence model,
     mesh method, BCs, endTime, outputs/metrics.
   * If underspecified: assume sensible defaults and list them explicitly before running.

2. Tutorial-first search (mandatory)
   * Identify 1-3 closest OpenFOAM tutorial cases.
   * Select one primary baseline; do not invent tutorial names/paths.
   * Prefer baselines that match: solver class + turbulence + mesh method.

3. Copy baseline & minimal edits
   * Copy the baseline tutorial case into a new working directory.
   * Modify only what is needed (mesh, BCs/ICs, turbulence, numerics, controlDict).
   * Prefer tutorial patterns for values; derive numbers only when necessary and explain briefly.

4. Geometry & mesh (as required by the case)
   * If geometry requires STL: generate it via Python+VTK in the selected conda env, validate it,
     then proceed with meshing (snappyHexMesh, etc.).
   * Run mesh generation (blockMesh and/or snappyHexMesh) following the baseline's pipeline.

5. Execute pipeline (mandatory)
   Run the full pipeline in the correct order for the chosen baseline:
   * mesh generation (blockMesh and/or snappyHexMesh)
   * checkMesh (must pass with acceptable quality)
   * decomposePar (only if running in parallel)
   * run the solver to required endTime

6. Error-driven repair loop (mandatory)
   If any stage fails:
   a) read the log and identify the root cause (first error, not downstream noise)
   b) apply the smallest correct fix (minimal file edits)
   c) rerun from the correct stage (do not restart earlier stages unnecessarily)
   d) repeat until it runs to endTime

   * Maintain a concise Fix Log: {symptom, cause, change, verification}.

7. Validate completion (mandatory)
   * Confirm endTime reached (from solver output and/or logs).
   * List output time directories (e.g., `ls -1 | grep -E '^[0-9]'` or case-specific path).
   * If required by the user: summarize key fields/metrics and where they are written.

# Deliverables (always output)
A) Tutorial baseline chosen (and 1-2 alternates) + rationale.
B) Exact commands executed, in order, and which logs to check.
C) Proof of completion: confirm endTime reached and list output time directories.
D) Minimal change summary vs baseline (files touched + key edits).
E) Fix Log (if any repairs were needed).
F) If STL used: STL generation method + validation summary.

# Defaults (use unless user overrides)
* Prefer the simplest solver matching physics; start from robust tutorial numerics.
* Transient: control Courant (or acoustic CFL for compressible); keep conservative schemes first.
* Steady: conservative relaxation and bounded schemes first.
* For internal flows with possible backflow: use tutorial-proven outlet BC patterns.

# OpenFOAM environment setup (mandatory)
* Source the OpenFOAM environment from the provided `<openfoam_root>` and chosen platform.
* Use tutorial cases as the baseline; do not create cases from scratch unless no tutorial fits.
\end{lstlisting}
\end{tcolorbox}\label{sec:agent-prompt}

\section{Descriptions for FoamBench-Advanced Planar 2D Obstacle-Flow Cases}\label{sec:foambench-cases}
\begin{tcolorbox}[colframe=black, colback=lightgray, coltitle=white, sharp corners=south, title=\texttt{\textbf{Cylinder\_LES}}]
Do an offset-cylinder simulation using the dynamicKEqn turbulence model and the PIMPLE algorithm. The computational domain is a 2D rectangle with x from -5 to 5 and y from -1.5 to 2.5. Place a circular cylinder of radius 0.3 centered at (-2, 1.2, 0). Use a fixed inlet velocity of 1 m/s at the left boundary, fixed value pressure at the outlet (right boundary), and no-slip wall conditions on all walls, including the cylinder surface. Use a time step of 0.0025 and write output every 0.2. Set final time to 2. Use a constant-viscosity model with nu = 0.01. Generate the geometry and mesh through multiblock hex meshing using only blockMeshDict. Use sufficient number of cells for refinement near the cylinder.
\end{tcolorbox}

\vspace{10pt}

\begin{tcolorbox}[colframe=black, colback=lightgray, coltitle=white, sharp corners=south, title=\texttt{\textbf{Diamond\_Obstacle\_KOMEGASST}}]
Perform an incompressible turbulent flow simulation over a 2D diamond obstacle using the kOmegaSST RANS turbulence model and pimpleFoam solver. The computational domain spans 0 to 15 in x direction and 0 to 5 in y direction and -0.5 to 0.5 in z direction. The diamond obstacle is a square rotated by 45 degrees with diagonal length of 1 unit centered at 2.5 x 2.5 x 0.0. Use one cell in z direction making the geometry effectively 2D. Refine the mesh near to diamond. Use sufficient grid points to discretize the domain. The left boundary is the inlet which uses a uniform velocity of (10,0,0) m/s. The right boundary is the outlet. Top and bottom boundaries are fixed walls with nop-slip condition. The front and back faces are empty. The diamond obstacle also has no-slip boundary condition on its surface. The kinematic viscosity is 2e-6 m2/s. Run till a final time of 10 s. Write the results at every 2 s. Use a maximum courant number of 1.0. Generate the geometry and mesh through multiblock hex meshing using only blockMeshDict.
\end{tcolorbox}

\vspace{10pt}

\begin{tcolorbox}[colframe=black, colback=lightgray, coltitle=white, sharp corners=south, title=\texttt{\textbf{Double\_Square\_SA}}]
Perform an incompressible turbulent flow simulation over two square obstacle using the SpalartAllmaras turbulence model and pimpleFoam solver. The computational domain spans 0 to 5 in x direction and 0 to 2.5 in y direction and -0.5 to 0.5 in z direction. One of the square obstacle is of size  1 unit x 1 unit x 1 unit centered at 1.5 x 1.25 x 0.0 and the other square obstacle is of size 1 unit x 1 unit x 1 unit centered at 3.5 x 1.25 x 0.0. Use one cell in z direction making the geometry effectively 2D. The left boundary is the inlet which uses a uniform velocity of (1,0,0) m/s. The right boundary is the outlet using fixed value pressure condition. Top and bottom boundaries are fixed walls with nop-slip condition. The front and back faces are empty. The rectangular obstacle also has no-slip boundary condition on its surface. The kinematic viscosity is 2e-6 m2/s. Run till a final time of 5 s. Write the results at every 0.5 s. Use a maximum courant number of 1.0. Generate the geometry and mesh through multiblock hex meshing using only blockMeshDict. Use sufficient number of cells for refinement near the obstacles.
\end{tcolorbox}

\vspace{10pt}

\begin{tcolorbox}[colframe=black, colback=lightgray, coltitle=white, sharp corners=south, title=\texttt{\textbf{Rectangular\_Obstacle\_KOMEGASST}}]
Perform an incompressible turbulent flow simulation over a 2D rectangular obstacle using the kOmegaSST RANS turbulence model and pimpleFoam solver. The computational domain spans 0 to 15 in x direction and 0 to 5 in y direction and -0.5 to 0.5 in z direction. The rectangular obstacle is of size  1 unit x 1 unit x 1 unit centered at 2.5 x 2.5 x 0.0. Use one cell in z direction making the geometry effectively 2D. The left boundary is the inlet which uses a uniform velocity of (1,0,0) m/s. The right boundary is the outlet using fixed value pressure condition. Top and bottom boundaries are fixed walls with nop-slip condition. The front and back faces are empty. The rectangular obstacle also has no-slip boundary condition on its surface. The kinematic viscosity is 2e-6 m2/s. Run till a final time of 5 s. Write the results at every 0.5 s. Use a maximum courant number of 1.0. Generate the geometry and mesh through multiblock hex meshing using only blockMeshDict. Use sufficient number of cells for refinement near the obstacles.
\end{tcolorbox}

\bibliographystyle{elsarticle-num}
\bibliography{ref}

\end{document}